\documentclass[aps,prl,amsfonts,reprint,superscriptaddress,footinbib]{revtex4-1}

\usepackage{graphicx}
\usepackage{amsfonts, amsmath, amstext, amssymb, amsfonts, amsxtra}
\usepackage{braket}
\usepackage{bbm}
\usepackage{bbold}
\usepackage[protrusion=true,expansion=true]{microtype}
\usepackage{wasysym}
\usepackage{cleveref}
\usepackage{color}

\newcommand{\id}{\mathbb{1}}  

\definecolor{myblue}{rgb}{0,0,0.75}

\newcommand{\Hs}{H}

\begin{document}

\title{Signatures of many-body localization in steady states of open quantum systems}

\author{I. Vakulchyk}
\affiliation{Center for Theoretical Physics of Complex Systems, IBS, Daejeon 34051, Republic of Korea}
\affiliation{Basic Science Program, Korea University of Science and Technology (UST), Daejeon 34113, Republic of Korea}
\author{I. Yusipov}
\affiliation{Department of Applied Mathematics, Lobachevsky  University, Nizhny Novgorod, 603950, Russia}
\author{M. Ivanchenko}
\affiliation{Department of Applied Mathematics, Lobachevsky  University, Nizhny Novgorod, 603950, Russia}
\author{S. Flach}
\affiliation{Center for Theoretical Physics of Complex Systems, IBS, Daejeon 34051, Republic of Korea}
\author{S. Denisov}
\affiliation{Institute of Physics, University of Augsburg, Universit\"{a}tsstra{\ss}e 1, 86159 Augsburg, Germany}
\affiliation{Department of Applied Mathematics, Lobachevsky  University, Nizhny Novgorod, 603950, Russia}



\begin{abstract}
Many-body localization (MBL) is a result of the balance between interference-based Anderson localization
and many-body interactions in an ultra-high dimensional Fock space.
It is usually expected that dissipation is blurring interference and destroying that balance so that the asymptotic state
of a system with an MBL Hamiltonian does not bear localization signatures. We demonstrate, within the framework of the Lindblad formalism,
that the system can be brought into a steady state with non-vanishing MBL signatures. 
We use a set of  dissipative operators 
acting on pairs of connected sites (or spins), and show that the difference between ergodic and MBL Hamiltonians 
is encoded in the  imbalance, entanglement entropy, and level spacing characteristics of the density operator. 
An MBL system
which is exposed to the combined impact of local dephasing and
pairwise dissipation evinces localization signatures hitherto absent in the dephasing-outshped steady state.
\end{abstract}


\maketitle

Many-body localization (MBL) is an extension of Anderson localization \cite{Anderson} into the world of many-body  systems \cite{MBL1,MBL2}. 
There is a spectrum of definitions/quantifiers of this multi-faceted phenomenon aimed to highlight peculiar properties of MBL systems, e.g. the
absence of conductivity \cite{MBL2} (even in the infinite temperature limit \cite{MBL1}), slow logarithmic growth of  the entanglement entropy 
after an interaction quench \cite{fazio2006,prosen2008,bardarson2012,serbyn2013a}, the existence of an extensive set of local integrals of 
motion \cite{serbyn2013b}, and specific spectral properties of MBL Hamiltonians \cite{oganesyan2007,serbyn2016}. There is a class of quantifiers which address 
properties of a single (eigen)state of an MBL system such as  short-range correlations \cite{Pal2010}, low spatial entanglement entropy \cite{nayak2013,s2014,khemani2017} and  
large spatial fluctuations of local observables \cite{bard2015}. 

Recently MBL became the subject of experiments with ultra-cold atoms \cite{exp1,Smith2016}. 
One of the important  questions concerns the impact of interactions with the environment
and the fate of MBL on the large time scales.
This question has been addressed recently in a series of papers \cite{dis1,dis2,dis3}, 
where the action of the environment was 
modeled with a Lindblad master equation and  a set of local dephasing operators. The answer confirmed intuition: Dissipation eventually 
destroys localization -- the steady state density operator  is the normalized identity -- but on the way to this state systems with MBL and non-MBL Hamiltonians behave 
notably differently (e.g., stretched  exponential vs exponential relaxations of some observables) \cite{foot1}.

Can we distinguish between MBL and non-MBL (ergodic) Hamiltonians by inspecting \textit{steady} states of the corresponding systems
when they are subjected to some  \textit{physically relevant} dissipation? It was recently  realized that dissipation is a full-fledged generator of evolution, no less complex 
and diverse than the unitary evolution generated by  Hamiltonians \cite{eng1,eng2,eng3};
e.g., dissipative mechanisms can be used to drive many-body systems into highly entangled pure states \cite{eng1}.

In this Letter we show that a controllable dissipation, when applied to a system with an MBL Hamiltonian, 
can sculpt an asymptotic state which bears detectable signatures of localization. 
These signatures can be revealed by using the population imbalance \cite{dis1,dis2,dis3} (a quantity measured in  
experiments \cite{exp1,exp2}), the operator spatial
entanglement entropy \cite{Prosen2005,dis3} (a generalization of the pure state spatial entanglement entropy to open systems), and the mean spectrum gap ratio \cite{oganesyan2007} 
of the steady state density operator.

\textit{Model.}  We study a conventional MBL model, an open-ended chain of $N$ (an even number) sites
occupied by $N/2$ spinless fermions. The fermions interact when occupying neighboring sites and are subject to
a random on-site potential $h_l$, $l=1,\dots,N$. The model Hamiltonian has the form 
\begin{equation}
 \Hs=-J\sum_{l=1}^{N}\left(c^\dagger_l c_{l+1} + c^\dagger_{l+1} c_{l}\right) + U\sum_{l=1}^N n_l n_{l+1}+\sum_{l=1}^Nh_l n_l,
 \label{eq:defHamiltonian}
\end{equation}
where $c^\dagger_l$ ($c_l$) creates (annihilates)  a fermion at site $l$, and $n_l=c^\dagger_l c_l$ is the local particle number operator.
Values $h_l$ are drawn from an uncorrelated uniform distribution on the interval $\left[-h,h\right]$. 
For $J=U=1$ (our choice here) this system undergoes a  many-body localization transition when $h >  h_{\mathrm{MBL}} \backsimeq 3.6$~\cite{Pal2010}. 
By using the Jordan-Wigner transformation, the system can be mapped onto a model of $N$ spins confined to the manifold $S^z = \sum_{l=1}^N s_l^z = 0$ \cite{book}.
This relation allows us to implement the time-evolving block decimation (TEBD) scheme generalized to matrix product operators \cite{Vidal} and propagate
the model system to its steady state. As the initial state we use $\varrho(0) = |\psi_0\rangle\langle\psi_0|$, $|\psi_0\rangle = |1010...10\rangle$.

\begin{figure*}[t]
\includegraphics[angle=0,width=1.99\columnwidth]{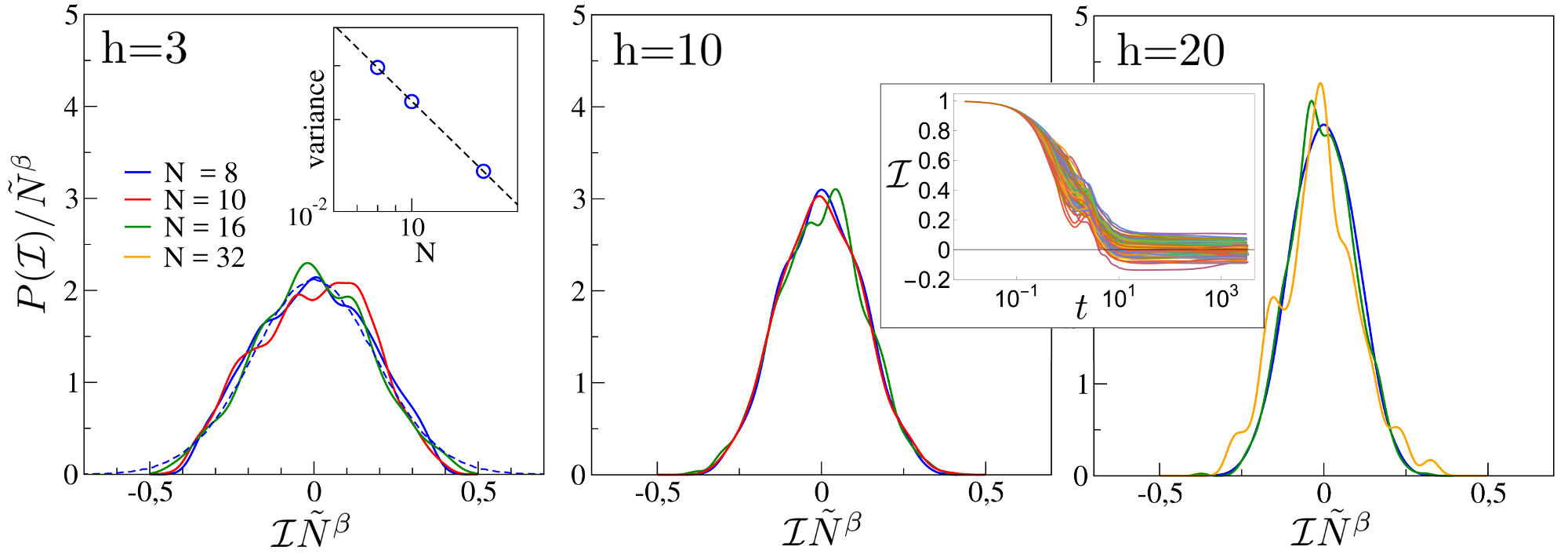}
\caption{Probability density function  $P(\mathcal{I})$ of the steady state imbalance $\mathcal{I}$ for different disorder strengths and system sizes.
Dashed line on panel (a) is the distribution sampled with the conditioned random $N$-partition of the unit  interval (see text)
Distributions are scaled with $\tilde{N}^{\beta}$, where $ \tilde{N} = N/8$ and exponent $\beta$ has values $0.55$ (a) and $0.8$ (b,c).
Insets: (a) scaling of the distribution variance with $N$ for $h=3$ (dashed line is the power-law $N^{2\beta}$) and (b-c) the time evolution 
of the imbalance for $10^2$ disorder realizations, $h=20$ and $N=32$,  obtained with the TEBD propagation \cite{Vidal}.
The parameters of the system, Eqs.~(\ref{eq:defHamiltonian} - \ref{eq:dissipator}), are $\gamma=0.1$,  $U=J=1$. 
Numbers $M$ of realizations are $10^4$ ($N=8$), $2 \cdot 10^3$ ($N=10$), $2 \cdot 10^3$ ($N=12,16$),
and $500$ ($N=32$).} \label{fig1}
\end{figure*}

The dissipation is captured with a master equation \cite{petr},
\begin{eqnarray} \nonumber
 \dot{\varrho}(t)= \mathcal{L}\varrho(t) = -i\left[\Hs,\varrho(t)\right]+ ~~~~~~~~~~~~~~~~~~~~~~~~~~~~~~~~~~~~~~~~~~\\ 
~~~~~ \sum_{s=1}^M \gamma_s \left[A_{s}\varrho(t)A^\dagger_{s}-\frac{1}{2}\{A^\dagger_{s}A_{s},\varrho(t)\}\right],
 \label{eq:defMaster}
\end{eqnarray}
where $\varrho(t)$ is the system density operator, and $A_{s}$ is the jump operator 
mimicking  the $s$-th dissipative channel of the environment, with rate $\gamma_s$. 

For Hermitian operators $A^{\dagger}_{s} = A_{s}$, the steady state density operator $\varrho_{\infty}$,  $\mathcal{L}\varrho_{\infty} = \mathbb{0}$, is the 
normalized identity $\varrho_{\infty} = \id/L$, $L =  \binom{N}{N/2}$. Hermitian dissipators grind any system into the infinite temperature state,
independently of the properties of system's Hamiltonian. This is  the case of local dephasing, $A_{l} = c^\dagger_l c_l$, $l=1,...,N$, considered in Refs.~\cite{dis1,dis2,dis3}. 
All other single-site operators (except for the identity, which does not influence the dynamics) 
do not preserve the evolution within the sector with fixed total number of particles $N/2$ ($S^z = 0$) \cite{alternative}.
On the other side, formally one could construct a non-Hermitian operator $A^i$ such that
$A^i|\phi_i\rangle = 0$, where $|\phi_i\rangle$ is the $i$-th eigenstate of the Hamiltonian $H$. Then 
the asymptotic state is  $\varrho_{\infty} = |\phi_i\rangle\langle \phi_i|$ \cite{eng1,eng2}.
However, such dissipators are too exotic and disorder-specific to be practically relevant.

We choose non-local dissipative operators
which act on a pair of neighboring sites \cite{eng1},
\begin{equation}
A_l=(c_{l}^{\dagger} + c_{l+1}^{\dagger})(c_{l}-c_{l+1}),~~~~\forall \gamma_l = \gamma.
\label{eq:dissipator}
\end{equation}
A physical interpretation of such dissipation is a 
chain coupled to a superfluid, which serves as a bath of Bogoliubov excitations; Raman transitions couple an antisymmetric state, by  the operator $(c_{l}-c_{l+1})$,
to the  excitations which then decay into a symmetric state, through the action of $(c_{l}^{\dagger} + c_{l+1}^{\dagger})$ \cite{eng1}.
With periodic boundary conditions and in the absence of disorder $h=0$  and interaction, $U=0$, these dissipators drive the system
into a uniform condensate (a dark state of all dissipators). 
For open boundary conditions and in the presence of the interactions and  disorder, the condensate is no longer an eigenstate of the Hamiltonian so
that the asymptotic state $\varrho_{\infty}$ is not pure and not homogeneous in general. 

To reveal the difference between  MBL ($h >  h_{\mathrm{MBL}}$) and ergodic ($h < h_{\mathrm{MBL}}$) Hamiltonians (\ref{eq:defHamiltonian}),
we calculate three quantifiers of $\varrho_{\infty}$.
We do this either  (i) by numerically finding $\varrho_{\infty}$ as a kernel of the Lindblad generator $\mathcal{L}$ \cite{kernel} ($N \leq 10$),
or (ii) by propagating   the matrix product representation of $\varrho(t)$,
until the quantifiers saturate to their asymptotic values ($10 \leq N \leq 32$) \cite{Vidal}. Note that our aim here is not to explore 
all possible regimes and parameter dependencies but to present a `proof of concept'. Therefore, for the following consideration we set $\gamma = 0.1$.


\begin{figure*}[t]
\includegraphics[angle=0,width=1.99\columnwidth]{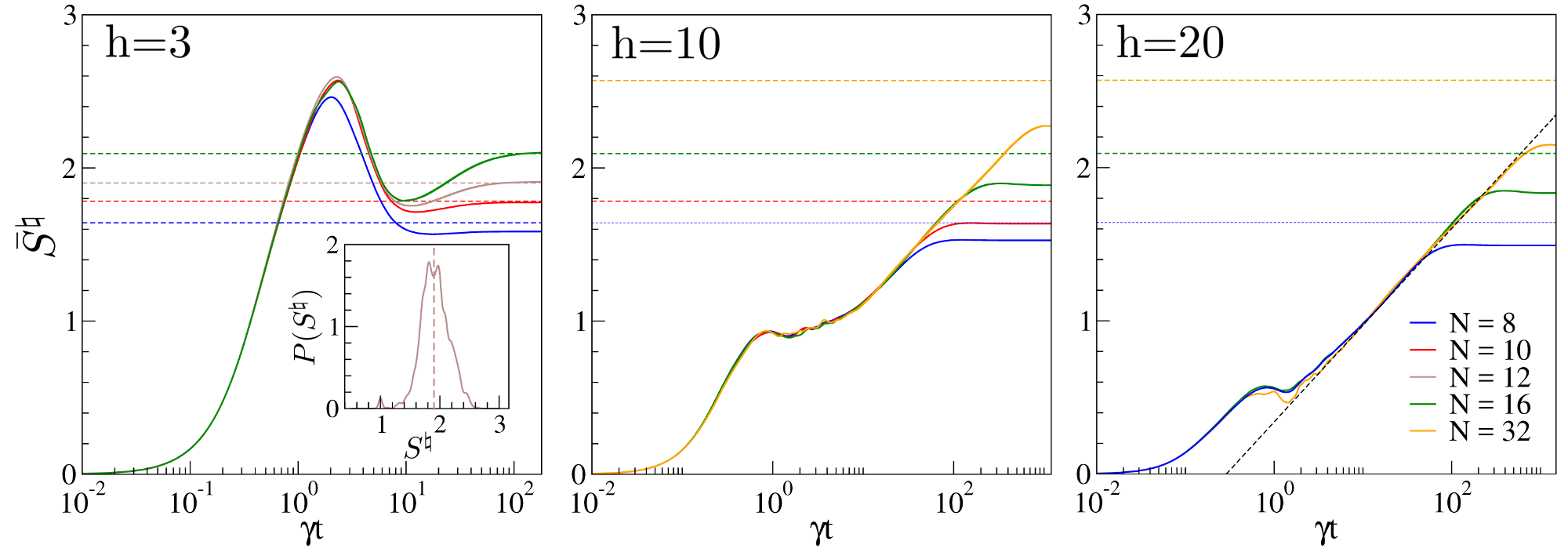}
\caption{Averaged operator-space entanglement entropy $\bar{S}^{\natural}(t)$ of the density operator $\varrho(t)$. Dashed lines are the values of the entropy 
for the maximally mixed (over the half-filled subspace) states \cite{znidaric2}. The dotted line on panel (c) is $\frac{1}{5}\log_2(t) + \mathrm{const}$.
Inset: The probability density function of the entropy of  individual disorder realizations, for $h=3$ and $N = 12$. Other parameters are as in Fig.~1.} \label{fig2}
\end{figure*}

\textit{Imbalance}. The imbalance is defined as
\begin{equation}
	\mathcal{I}(t) = \frac{N_o(t)-N_e(t)}{N/2},
 \label{eq:imbalance}
\end{equation}
where $N_{o}$ ($N_{e}$) is the number of fermions in odd (even) sites.
This characteristics was measured in the recent experiments to quantify the MBL \cite{exp1,exp2} (note that
due to particle loss time-dependent denominators  were used).

When dissipation is non-Hermitian, the asymptotic imbalance $\mathcal{I} = \lim_{t \rightarrow \infty} \mathcal{I}(t)$ is 
a real-valued random variable $\mathcal{I}_s$, different for different disorder realizations, $s \in \{1,2,...,M\}$. 
In the absence of any statistical theory of this quantity, we consider $\{\mathcal{I}_s\}$  as a set of independent and identically distributed (iid) random 
variables with a probability density function (pdf) $P(\mathcal{I})$. 
In the ergodic regime $h <  h_{\mathrm{MBL}}$, a configuration of site populations $n^s_l = Tr[\varrho_{\infty} n_l]$ can be modeled as a random $N$-partition of the unit interval
$\mathbf{x}^s = \{x^s_1,x^s_2,...,x^s_N\}$, $\sum_l x^s_l =1$, uniformly distributed over the subspace $\mathcal{A} =\{\mathbf{x}:\forall x_l \leq 2/N\}$ ('no more than one particle per site'). 
The sampling results for $\mathcal{I}_{\mathrm{mod}}[N] = \sum_{l=1}^N (-1)^lx_l$ are in a good agreement with 
the sampling of the  model for $h=3$ ~\cite{MPO}; see Fig.~\ref{fig1}(a). The only notable difference is in the tail regions:
While the stochastic pdf has unbounded tails, the pdf for the model (1-3) is always confined to the interval $[-1/2,1/2]$.

Being the sum of $N$ iid random variables, $\mathcal{I}[N]$ is subject to the Central Limit Theorem \cite{clt}.
Then the scaling $N^{-\beta}P(N^{\beta}\mathcal{I}[N])$ with $\beta=0.5$ is expected. The variance of the sampled pdf $P(\mathcal{I})$ 
yields the exponent $\beta \backsimeq 0.55$ in the ergodic regime, see inset in Fig.~\ref{fig1}(a). 
For large disorder we find $\beta \approx 0.8$, Figs.~\ref{fig1}(b-c), which indicates a transition into the MBL phase.
The narrowing of the pdf can be explained by the presence of short-range anti-correlations which tie neighboring sites,  a marked feature of MBL states  \cite{oganesyan2007}.

\textit{Operator-space entanglement entropy (OSSE)}. This quantity  was introduced by Prosen and Pi\v{z}orn~\cite{Prosen2007} as an operator generalization of 
the spatial entanglement entropy (defined for pure states). OSSE was implemented for the density operator in order to monitor the 
relaxation of an open MBL system to the infinite temperature state~\cite{dis3}. To calculate this quantity, 
one should split the chain into two (equal in our case) parts and calculate the Schmidt decomposition of the density operator, $\varrho = \sum_k\sqrt{\mu_k} C_k \otimes D_k$,
where the operators $C_k$  ($D_k$)  act non-trivially on the left (right) half only and form a complete Hilbert-Schmidt basis in the corresponding subspace. 
The normalized coefficients $\bar{\mu}_k$ define the entropy value $S^{\natural} = -\sum_k \bar{\mu}_k \log_2\bar{\mu}_k$. When the state is pure, $S^{\natural}$ is twice the standard 
entanglement entropy ~\cite{Znidaric2008}.

\begin{figure*}[t]
\includegraphics[angle=0,width=1.99\columnwidth]{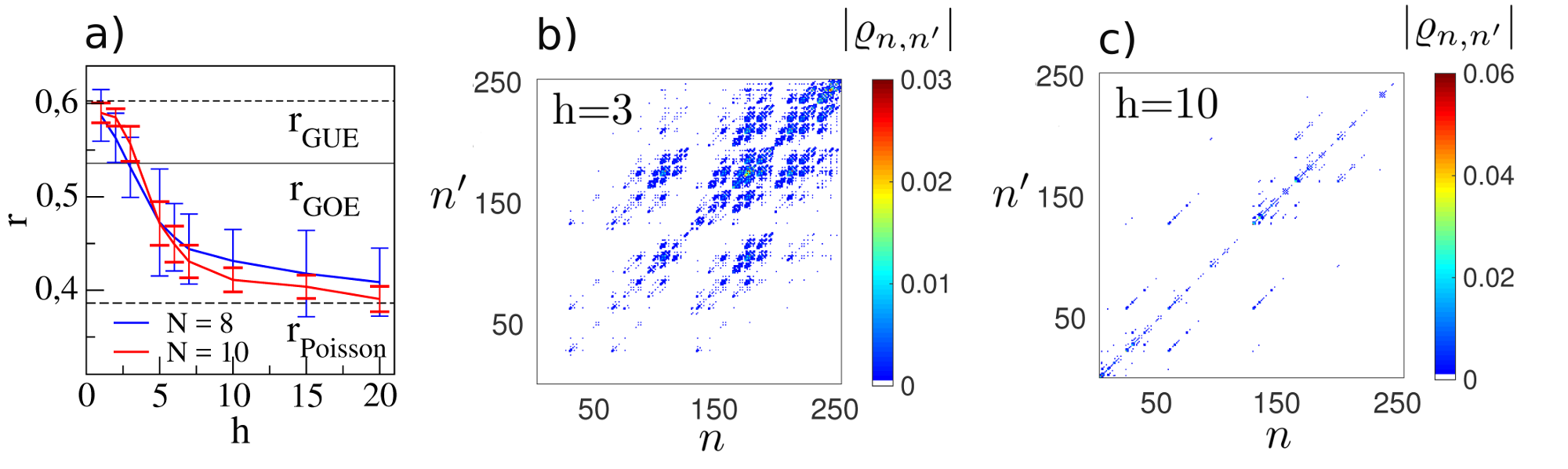}
\caption{ (a) Averaged ratio of consecutive level spacing $r$ of 
$\varrho_{\infty}$ as a function of disorder strength $h$. 
The ratio is sampled for chains
with  $N=8$  and $10$ sites and averaged (for every value of $h$) over $10^2$ disorder realizations. The error bars show the variance of the ratio averaged over the spectrum 
of a single disorder realization.
(b-c) Absolute values of the elements of the steady state density matrix for a single disorder realization and two different values of $h$. 
The matrices are expressed in the Fock basis (for the half-filling sector)
sorted in the lexicographical order. Only elements with absolute value larger than $10^{-5}$ are shown.
Other parameters are the same as in Fig.~1.} 
\label{fig3}
\end{figure*}

In the ergodic phase $h = 3$ we find that for $N \geq 10$ the averaged (over the disorder)  OSSE $\bar{S}^{\natural}(t)$  saturates to $S^{\natural}(\id_{\mathrm{HF}})$, which is
the entropy corresponding to the state maximally mixed over the half-filled subspace $\mathcal{H}_L$ \cite{znidaric2}, Fig.~\ref{fig2}(a). This implies
an \textit{effective} thermalization of the system: At variance to the case of local dephasing \cite{dis3}, the individual realization entropy values are not all identical 
to $S^{\natural}(\id_{\mathrm{HF}})$ but distributed around it, see inset in Fig.~\ref{fig2}(a) [as we show in the next section, a single ergodic steady state is far from being maximally mixed].
The initial short-time evolution of the entropy follows the Hamiltonian path. It is a linear growth, which in the absence 
of the dissipation will saturate to the Page value \cite{page}, $S^{\natural}_{\mathrm{Page}} \backsimeq N - 1$, corresponding to the entropy of a typical random pure state 
uniformly 'smeared' over $\mathcal{H}_L$. After  time
$t \gtrsim \gamma^{-1}$ the contribution of the dissipative part of the generator $\mathcal{L}$ starts to  the  dynamics becomes tangible and eventually brings the entropy down 
to an asymptotic value near
$S^{\natural}(\id_{\mathrm{HF}}) \ll S^{\natural}_{\mathrm{Page}}$. 

In the MBL phase, the averaged OSSE saturates to values below $S^{\natural}(\id_{\mathrm{HF}})$, see Figs.~\ref{fig2}(b-c). This can be explained by 
generalizing the argument used in Ref.~\cite{Pal2010}
for the Hamiltonian case. While in the ergodic phase all -- even distant -- sites (spins) are 'tied' by the conservation of the total particle number (total spin), 
in the MBL phase the correlations are short-ranged and restricted by the localization length.
Therefore, the entanglement entropy is lower in the MBL phase.  The relaxation dynamics of the OSSE in the strong localization limit is marked by a logarithmic 
growth,  $S^{\natural}(t) \backsimeq g \log_2(t)$, a feature
found before with local dephasing \cite{dis1,dis2,dis3}. The prefactor $g$, as conjectured in Ref.~\cite{dis3}, is related to the scaling 
of the spectral gap of the generator $\mathcal{L}$  with the size of the system. If the gap scales as $N^\nu$, then $g \backsimeq \frac{1}{2\nu}$.
For $h=20$ we find $g \backsimeq 0.2$, see Fig.~\ref{fig2}(c). This gives the value of the scaling exponent $\nu \backsimeq 2.5$. 
Whether this is a universal value belonging to one of the universality classes, discussed in Ref.~\cite{znidaric2}, is an interesting question which demands a more detailed analysis.

\textit{Ratio of consecutive level spacing for the steady state density operator}. There is a genetic link  between changes in the spectral statistics of many-body Hamiltonians 
and ergodic-MBL transitions \cite{oganesyan2007,serbyn2016}. According to the quantum chaos theory,  Poisson  and Wigner-Dyson distributions of the energy spacing $\delta_j = E_{j+1} - E_{j}$
correspond to  regular (integrable) and chaotic (non-integrable) quantum systems \cite{haake}. Similarly, we can expect Poisson and Wigner-Dyson
distributions for MBL and ergodic Hamiltonians, respectively \cite{serbyn2016}.
However, these indicators assume the uniform level density which is rarely the case with  physical Hamiltonians \cite{unfolding}. 
To circumvent this problem, Oganesyan and Huse considered the distribution of the ratios $r_j = \min[\lambda_j,\lambda_j^{-1}]$,  $\lambda_j = \delta_j/\delta_{j-1}$, which
do not depend on the local density of states \cite{oganesyan2007}. It follows that spectral averages of $r$ yield $r_{\mathrm{Poisson}} \backsimeq 0.386$
for Poisson random variables, $r_{\mathrm{GOE}} \backsimeq 0.536$ for Gaussian orthogonal  (GOE),   and $r_{\mathrm{GUE}} \backsimeq 0.603$ for
Gaussian unitary (GUE) ensembles~\cite{atas}.
The ergodic-to-MBL transitions correspond to the passage from $r_{\mathrm{Poisson}}$ to $r_{\mathrm{GOE}}$ \cite{oganesyan2007}.

In another context, Prosen and \v{Z}nidari\u{c} proposed to quantify the non-equilibrium steady state density operators in terms of their level spacing distributions \cite{Prosen2013}. 
They found that
the transition from integrability 
to non-integrability \cite{def_non} corresponds to the Poisson-to-\textit{GUE} transition in 
the distribution of the level spacing of the density operator. Here we follow this idea, but implement the averaged ratio of consecutive level spacing $r$ instead.

We find that in the ergodic phase the spectrum of the steady state density operator displays $r$ values close to $r_{\mathrm{GUE}}$, while 
in the limit  of strong localization its value approaches $r_{\mathrm{Poisson}}$,  see Fig.~\ref{fig3}(a). 
This correspondence improves with increasing $N$. 
The structure of the density matrices $\varrho_{\infty}$ is notably different in the ergodic and strong localization regimes, see Figs.~\ref{fig3}(b-c):
While in the ergodic phase matrices exhibit a well-developed off-diagonal structure and thus a relatively high purity and interference, in the deep MBL regime they 
have near diagonal structure, with a few `hot spots' (a similar structure was found before in the context of dissipative single-particle localization \cite{Yusipov2017}).

\textit{Pairwise dissipation on top of local dephasing.} Consider a Lindblad generator $\mathcal{L} = \mathcal{L}_{\mathrm{MBL}} + 
\gamma_{\mathrm{deph}} \mathcal{L}_{\mathrm{deph}} + \gamma_{\mathrm{pair}} \mathcal{L}_{\mathrm{pair}}$,
where $\mathcal{L}_{\mathrm{MBL}} =-i[H_{\mathrm{MBL}},\cdot]$ and the two next terms are dissipative Liouvillians corresponding to local 
dephasing and pairwise dissipation
respectively.
In the limit $\gamma_{\mathrm{pair}} = 0$, any whatever small but finite dephasing  $\gamma_{\mathrm{deph}} \neq 0$
will eventually bring the system into the  maximally mixed state with no MBL signatures.
Assume now that $\gamma_{\mathrm{deph}} \|\mathcal{L}_{\mathrm{deph}}\| \ll \|\mathcal{L}_{\mathrm{MBL}}\|$, where $\|\cdot\|$  is a
suitable operator norm \cite{norms} defined on the set of, e.g., matrix product operator (MPO) states \cite{Verstraete2004}, which serve  a proper basis for 
weakly-entangled mixed states. By adding pairwise dissipation
$ \gamma_{\mathrm{deph}} \|\mathcal{L}_{\mathrm{deph}}\|  < \gamma_{\mathrm{pair}} \|\mathcal{L}_{\mathrm{pair}}\|   \ll \|\mathcal{L}_{\mathrm{MBL}}\|$ it is
possible to create  a new steady state, with the corresponding density operator bears the signatures of the MBL (though to the degree dependent
on the relative values of  $\gamma_{\mathrm{deph}}$ and $\gamma_{\mathrm{pair}}$) \cite{check}. This conjecture is based on the stability of 
many-body dissipative systems with no faster than linear (in time) growth of the support of initially localized operators \cite{Cubitt2015}.

\textit{Discussion.} We proposed three quantitative identifiers of MBL in open systems. The imbalance statistics is accessible in experiments \cite{exp1,exp2}
but requires studying systems of different sizes. The operator-space entanglement entropy 
indicates differences between phases both in the asymptotic limit and during the relaxation towards it. The level spacing of the asymptotic density 
operator $\varrho_{\infty}$ bridges  MBL and quantum chaos theory \cite{haake, Prosen2013}.
The operator provides  complete information on propertis  of the system in the asymptotic limit (including values of all three identifiers); 
however, its numerical resolution is possible for relatively small systems, $N \leq 10$. The  TEBD propagation is useful in case we want to explore 
the relaxation of the system to its steady state. To address the steady state directly,
it is more advantageous to use recently developed variational methods \cite{Cui2015}. MBL steady states naturally fulfill the conditions imposed on the
matrix-product operators (MPOs) so that the MPO ansatz \cite{Verstraete2004} should work well in this case.

The considered regular pairwise dissipation is perhaps not the best choice to create an MBL steady state. 
Such dissipation tries to build a long-range entanglement  in the system \cite{eng1}, and in this sense it does not favor localization.
The states we observed are the result of the antagonistic competition between the unitary MBL dynamics and dissipation. 
However, this is  the only physically reasonable \cite{Diehl2012} type of non-Hermitian dissipation, preserving the number of particles, 
which we found in the recent literature. Future studies could consider the incorporation of the disorder into local rates $\gamma_l$. This idea leads to an intriguing question 
of creation
MBL  states by  dissipative means solely, without Hamiltonian disorder. Disordered pairwise dissipation  acquires relevance in the context of 
recent experiments with dissipatively coupled  exciton-polariton condensate arrays \cite{polyariton}.

\begin{acknowledgments}
Numerical simulations were performed on the PCS IBS cluster (Daejeon),  the Lobachevsky super-cluster (Nizhny Novgorod), and the MPIPKS cluster (Dresden). S. D., I.Y. and M. I. acknowledge support by the Russian Science Foundation via grant No. 15-12-20029. I.V. and S.F. acknowledge support by the Institute for Basic Science in Korea (IBS-R024-D1).
\end{acknowledgments}


\begin{thebibliography}{1000}

\bibitem{Anderson} P. W. Anderson Rev. Mod. Phys. \textbf{50}, 191 (1978).
\bibitem{MBL1} D. M. Basko, I. L. Aleiner, B. L. Altshuler, Ann. Phys. (Amsterdam) \textbf{321}, 1126 (2006).
\bibitem{MBL2} I. V. Gornyi, A. D. Mirlin, D. G. Polyakov, Phys. Rev. Lett. \textbf{95}, 206603 (2005).
\bibitem{fazio2006} G. De Chiara, S. Montangero, P. Calabrese, and R. Fazio, J. Stat. Mech. (2006) P03001.
\bibitem{prosen2008} M. \v{Z}nidari\u{c}, T. Prosen, and P. Prelov\v{s}ek, Phys. Rev. B 77, 064426 (2008).
\bibitem{bardarson2012} J. H. Bardarson, F. Pollmann, and J. E. Moore, Phys. Rev. Lett. \textbf{109}, 017202 (2012).
\bibitem{serbyn2013a} M. Serbyn, Z. Papi\'{c}, and D. A. Abanin, Phys. Rev. Lett. \textbf{110}, 260601 (2013).
\bibitem{serbyn2013b} M. Serbyn, Z. Papi\'{c}, and D. A. Abanin, Phys. Rev. Lett. \textbf{111}, 127201 (2013).
\bibitem{oganesyan2007} V. Oganesyan and D. A. Huse, Phys. Rev. B \textbf{75}, 155111 (2007).
\bibitem{serbyn2016} M. Serbyn and J. E. Moore, Phys. Rev. B \textbf{93}, 041424 (2016).
\bibitem{Pal2010} A.~Pal  and D.~A. Huse, Phys. Rev. B 82, 174411 (2010). 
\bibitem{nayak2013} B. Bauer and Ch. Nayak, J. Stat. Mech. (2013) P09005.
\bibitem{s2014} J. A. Kj\"{a}ll, J. H. Bardarson, and F. Pollmann, Phys. Rev. Lett. \textbf{113}, 107204 (2014).
\bibitem{khemani2017}   V. Khemani, S. P. Lim, D. N. Sheng, and David A. Huse, Phys. Rev. X \textbf{7}, 021013 (2017).
\bibitem{bard2015} S. Bera, H. Schomerus, F. Heidrich-Meisner, and J. H. Bardarson, Phys. Rev. Lett. \textbf{115}, 046603 (2015).
\bibitem{exp1} M. Schreiber \textit{et al.}, Science \textbf{349}, 842 (2015); P. Bordia, H. L\"{u}schen, U. Schneider, M. Knap, and I. Bloch,
Science \textbf{352}, 1547 (2016);  P. Bordia,	H. L\"{u}schen,	U. Schneider, M. Knap, and I. Bloch, Nature Phys. \textbf{13}, 460 (2017).
\bibitem{Smith2016} J. Smith, A. Lee, P. Richerme, B. Neyenhuis, P. W. Hess, P. Hauke, M. Heyl, D. A. Huse, and C. Monroe, Nature Phys. 12, 907 (2016).
\bibitem{dis1} E. Levi, M. Heyl, I. Lesanovsky, J. P. Garrahan, Phys. Rev. Lett. {\bf 116}, 237203 (2015).
\bibitem{dis2} M. F. Fisher, M. Maksymenko, E. Altman, Phys. Rev. Lett. {\bf 116}, 160401 (2016).
\bibitem{dis3} M. V. Medvedyeva, T. Prosen, M \v{Z}nidari\u{c}, Phys. Rev. B \textbf{93}, 094205 (2016).
\bibitem{foot1} This difference was observed in a very recent experiment \cite{exp2}.
\bibitem{eng1} S. Diehl, A. Micheli, A. Kantian, B. Kraus, H. P. B\"{u}chler, P. Zoller, Nature Physics 4, 878 (2008).
\bibitem{eng2} B. Kraus H. P. B\"{u}chler, S. Diehl, A. Kantian, A. Micheli, P. Zoller, Phys. Rev. A \textbf{78}, 042307 (2008).
\bibitem{eng3}  F. Verstraete, M. M. Wolf, and J. I. Cirac, Nature Phys. \textbf{5}, 633 (2009).
\bibitem{exp2} H. P. L\"{u}schen et al., Phys. Rev. X \textbf{7}, 011034 (2017).
\bibitem{Prosen2005} T. Prosen and I. Pi\v{z}orn, Phys. Rev. A \textbf{72}, 032317 (2005); 
\bibitem{book} P. Coleman, \textit{Introduction to Many-Body Physics} (Cambridge University Press, 2015).
\bibitem{Vidal} M. Zwolak and G. Vidal, Phys. Rev. Lett. 93, 207205 (2004); R. Or\'{u}s and G. Vidal, Phys. Rev. B 78, 155117 (2008).
\bibitem{petr} H.-P. Breuer  and F. Petruccione, \textit{Theory of Open Quantum Systems} (Oxford University Press, 2002).
\bibitem{alternative} In Ref.~\cite{dis2} dissipation in the form of a single-particle loss operator, $A_{l}=c_l$, was  considered.
Evidently, the steady state in this case is the vacuum $|0^N\rangle$.
\bibitem{kernel} We vectorize the density  operator in the basis
of the generalized Gell-Mann matrices \cite{Alicki} and then solve the obtained real-valued system of $L^2-1$ linear equations.
\bibitem{Alicki} R. Alicki and K. Lendi, \textit{Quantum Dynamical Semigroups and Applications}, Lecture Notes in Physics Vol. 286 (Springer, Berlin, 1987).
\bibitem{MPO} We are not going  deep into the ergodic phase in order to be able to use the TEBD propagation with reasonable low bond dimension ($ \leq 500$)
and high accuracy. For $N=10$ and $10^2$ individual disorder realizations, we compare results of the numerically exact spectral propagation (by diagonalizing generator $\mathcal{L}$
and implementing its dual basis and eigenvalues) and of the TEBD propagation; the relative error for the imbalance and entropy did not exceed $10^{-4}$ for 
all three values of $h$.
\bibitem{clt} P. Billingsley, \textit{Probability and Measure} (John Wiley $\&$ Sons, 1995). 
\bibitem{Prosen2007} T. Prosen and I. Pi\v{z}orn, Phys. Rev. A. \textbf{76}, 032316 (2007).
\bibitem{Znidaric2008} M. \v{Z}nidari\u{c}, T. Prosen and I. Pi\v{z}orn, Phys. Rev. A \textbf{78}, 022103 (2008).
\bibitem{znidaric2} M. \v{Z}nidari\u{c}, Phys. Rev. E \textbf{92}, 042143 (2015).
\bibitem{page} D. N. Page, Phys. Rev. Lett. \textbf{71}, 1291 (1993).
\bibitem{haake}  F. Haake, \textit{Quantum Signatures of Chaos} (Springer, Berlin-Heidelberg, 2013).
\bibitem{unfolding} In order to relate spectra of these Hamiltonians to a specific universality  class, 
a so-called unfolding transformation has to be performed; see  C. E. Porter, \textit{Statisitical Theories of Spectra: Fluctuations} (Academic Press, New York, 1965).
\bibitem{atas} Y. Y. Atas, E. Bogomolny. O. Giraud, and G. Roux, Phys. Rev. Lett. \textbf{110}, 084101 (2013).
\bibitem{Prosen2013} T. Prosen and M. \v{Z}nidari\u{c}, Phys. Rev. Lett. \textbf{111}, 124101 (2013).
\bibitem{def_non} The integrability of  density operators was defined in Ref.~\cite{Prosen2013} as 
'the existence of an algebraic procedure for their construction in finitely many steps', e.g.,  by using the matrix produc state 
ansatz [see  M. \v{Z}nidari\u{c}, J. Phys. A \textbf{43}, 415004 (2010)].
\bibitem{Luitz2015} D. J. Luitz, N. Laflorencie, and F. Alet, Phys. Rev. B \textbf{91}, 081103(R) (2015).
\bibitem{Yusipov2017} I. Yusipov, T. Laptyeva, S. Denisov, and M. Ivanchenko, Phys. Rev. Lett. 118, 070402 (2017).
\bibitem{norms} Since all norms defined for bounded operators on a finite dimensional space are equivalent [J. B. Conway,\textit{ A Course in Functional Analysis} (Springer, NY, 1990)], 
the particular norm choice is not important.
\bibitem{Verstraete2004} F. Verstraete, J. J. Garcıa-Ripoll, and J. I. Cirac, Phys. Rev. Lett. 93, 207204 (2004).
\bibitem{check} We verified this conjecture in simulations (results are not presented here).
\bibitem{Cubitt2015} A. Lucia, T. S. Cubitt, S. Michalakis, D. P\'{e}rez-Garc\'{\i}a, Phys. Rev. A \textbf{91}, 040302 (2015); 
T. S. Cubitt, A. Lucia, S. Michalakis, D. P\'{e}rez-Garc\'{\i}a, Comm.  Math. Phys. \textbf{337}, 1275 (2015).
\bibitem{Cui2015} J. Cui, J. Cirac, M. C. Ba\~{n}uls, Phys. Rev. Lett. \textbf{114}, 220601 (2015).
\bibitem{Diehl2012} D. Marcos, A. Tomadin, S. Diehl, and P. Rabl, New J. Phys. \textbf{15}, 055005 (2012).
\bibitem{polyariton} S. R. K. Rodriguez, A. Amo, I. Sagnes, L. Le Gratiet, E. Galopin, A. Lemaitre, and J. Bloch, Nature Comm. \textbf{7},  11887 (2016).




\end{thebibliography}
\end{document}